\documentclass[aps,prb,twocolumn,superscriptaddress]{revtex4-2}

\usepackage[utf8]{inputenc}
\usepackage[english]{babel}
\usepackage{mathtools}
\usepackage{physics}
\usepackage{braket}
\usepackage{bbm}
\usepackage{bm}
\usepackage{amsbsy}
\usepackage{amsthm}
\usepackage{amssymb}
\usepackage{amsfonts}
\usepackage{amsmath}
\usepackage{dsfont} 
\usepackage{graphicx, caption} 
\usepackage{tikz}
\usepackage{adjustbox}
\usetikzlibrary{positioning}
\usepackage{subcaption}
\usepackage{epsfig}
\usepackage{epstopdf}
\usepackage{dsfont}
\usepackage{multibib}
\usepackage{xcolor}
\usepackage{color}
\usepackage{verbatim}
\usepackage{nomencl}
\usepackage[colorlinks]{hyperref}
\usepackage{float}
\usepackage{graphicx}
\usepackage{overpic}
\usepackage[normalem]{ulem}

\begin{document}

	\title{Sensing ac fields with quantum many-body scars}

         \author{Matheus Fibger}
	\affiliation{Instituto de F\'isica, Universidade Federal Fluminense, Av. Gal. Milton Tavares de Souza s/n, Gragoat\'a, 24210-346 Niter\'oi, Rio de Janeiro, Brazil}
	\author{Andrei Tsypilnikov}
    \affiliation{Instituto de F\'isica, Universidade Federal Fluminense, Av. Gal. Milton Tavares de Souza s/n, Gragoat\'a, 24210-346 Niter\'oi, Rio de Janeiro, Brazil}
    \author{Thiago R. de Oliveira}
    \affiliation{Instituto de F\'isica, Universidade Federal Fluminense, Av. Gal. Milton Tavares de Souza s/n, Gragoat\'a, 24210-346 Niter\'oi, Rio de Janeiro, Brazil}
    \author{Fernando Iemini}
	\affiliation{Instituto de F\'isica, Universidade Federal Fluminense, Av. Gal. Milton Tavares de Souza s/n, Gragoat\'a, 24210-346 Niter\'oi, Rio de Janeiro, Brazil}

\begin{abstract}

Quantum many-body scars (MBS) exhibit weak ergodicity breaking and long-lived coherent dynamics within an otherwise thermal spectrum. We investigate their metrological properties using the quantum Fisher information (QFI), focusing on estimating the amplitude of a weak AC field in the PXP model. We show that the approximately uniform energy spacing of the scar tower enables collective resonant processes when the driving frequency matches integer multiples of the scar gap, resulting in a quadratic-in-time growth of the QFI over an extended time window. We analyze how the connectivity induced by different probe operators shapes sensing performance and demonstrate that staggered magnetization leads to a more favorable growth of the QFI with system size than homogeneous magnetization. Through frequency scanning and finite-size analysis, we characterize the scaling of the QFI with the number of particles. Finally, we develop a single-tower approximation under resonant driving, deriving a compact analytical expression that captures the time dependence and system-size scaling of the QFI. Our results establish how to leverage structured non-ergodic dynamics in quantum sensing protocols.

\end{abstract}

	\pacs{}
    
	\maketitle

\section{Introduction}

Quantum metrology explores how to exploit quantum systems to enhance the precision of parameter estimation beyond classical limits \cite{giovannetti2004quantum,giovannetti2011advances}. By leveraging uniquely quantum resources such as coherence, entanglement, and many-body correlations, it is possible to achieve sensitivities that surpass the standard quantum limit and approach the Heisenberg limit \cite{giovannetti2011advances,pezze2018quantum}.  However, in realistic many-body systems, the generation and preservation of such resources are typically hindered by decoherence and thermalization, which restrict the available interrogation time and degrade metrological performance \cite{huelga1997improvement,escher2011general}.

A potential route towards enhanced sensing has recently emerged from the study of non-equilibrium quantum dynamics in isolated many-body systems \cite{montenegro2025quantum}. In particular, systems that exhibit weak ergodicity breaking can sustain long-lived coherent dynamics even in the absence of fine-tuning \cite{nandkishore2015manybody,abanin2019colloquium}. Such persistent coherence can extend the effective interrogation time and preserve quantum correlations, thereby enhancing the achievable precision in parameter estimation. A well-established example is quantum many-body scars, which correspond to atypical, non-thermal eigenstates embedded within an otherwise thermal spectrum \cite{serbyn2021quantum}. These states give rise to coherent oscillatory dynamics and periodic revivals for specific initial conditions, defying the expectations of the eigenstate thermalization hypothesis \cite{deutsch1991quantum,srednicki1994chaos,rigol2008thermalization}.

The presence of long-lived coherence and structured dynamics in scarred systems suggests they may serve as a useful resource for quantum sensing. Unlike highly entangled states engineered for optimal metrological performance, which are often fragile, scarred dynamics arise naturally from the system’s Hamiltonian and can persist over extended time scales. 
The use of quantum many-body scars for quantum sensing of DC fields has been previously investigated in Refs.~\cite{dooley2023entanglement,dooley2021robust}, highlighting their potential for metrological applications based on coherent dynamics and quantum Fisher information. Moreover, coherent scar behavior has been experimentally observed in Rydberg atom arrays described by the PXP model, where coherent revivals and slow thermalization have been reported \cite{bernien2017probing,turner2018weak}.
This raises the question of whether one can harness such non-ergodic behavior to improve parameter estimation protocols, particularly in regimes where coherence time is the primary limiting factor. In this work, we address this question and investigate the metrological properties of quantum many-body scar systems by focusing on the estimation of a weak time-dependent field. We consider the paradigmatic PXP model, which hosts a well-defined tower of scarred eigenstates with approximately uniform energy spacing \cite{turner2018weak,serbyn2021quantum}. Using the QFI as a figure of merit \cite{paris2009quantum}, we analyze how the structure of the scar spectrum and the choice of probe operators influence the sensing performance. In particular, we show that resonant driving can activate collective transitions across the scar tower, leading to enhanced, sustained growth of the QFI. 

Furthermore, we examine how different operator structures affect the connectivity within the scar manifold and, consequently, the scaling of the metrological gain with system size. In this work, we use two types of external fields whose amplitudes we aim to measure: the total magnetization and staggered magnetization. For the total magnetization, the scar eigenstates connect more favorably with the next-nearest neighbor; in the staggered case, they connect better with the nearest neighbor; and in both cases, they scale superlinearly with system size. 
Moreover, by combining numerical analysis with an analytical approximation based on a single scar tower, we provide insight into the mechanisms underlying the observed enhancement. Our results demonstrate that structured non-ergodic dynamics can be directly converted into metrological advantage, offering a complementary strategy to conventional approaches based on highly entangled states.

This paper is organized as follows. In Sec .~II, we review the basic concepts of quantum parameter estimation and introduce the Fisher information framework. In Sec.~III, we present the model and discuss the relevant properties of many-body scars. In Sec.~IV, we analyze the sensing performance, including resonant effects and finite-size scaling. In Sec.~V, we develop an analytical description based on a single-tower approximation. Finally, we summarize our results and discuss future directions in Sec.~VI.

\section{Fisher Information}

In this section, we review the main concepts of metrology and estimation theory. In a few words, the goal of metrology is to determine the value of an unknown parameter $h$ from empirical observations, a process formalized by estimation theory.
The efficacy of this process (that is, the precision of the resulting estimate) is not arbitrary but fundamentally governed by the Fisher information (FI), as we will discuss.

\begin{figure}
  \centering
  \vspace{0.5cm}
  \adjustbox{max width=\columnwidth}{
    \begin{tikzpicture}[every node/.style={rectangle, draw, minimum height=2cm, minimum width=1.5cm, text centered, scale=1}]
      \node (A) [fill=red!30, align=center] {State \\ preparation\\ $\hat{\rho}_0$};
      \node (B) [fill=green!30, right=0.5cm of A, align=center] {Encoding \\ information \\ $\hat \rho(t|h)$};
      \node (C) [fill=blue!30, right=0.5cm of B, align=center] {Readout \\ outcomes \\ $p(\mu_i|h)$};
      \node (D) [fill=yellow!30, right=0.5cm of C, align=center] {Estimation \\$\Theta[\mu]$\\ $(\Delta h)^2 = (\Theta[\mu]-h)^2$};
      \draw[->] (A) -- (B);
      \draw[->] (B) -- (C);
      \draw[->] (C) -- (D);
    \end{tikzpicture}
  }
  \caption{Fundamental phases of a quantum sensing protocol: (i) prepare a probe state, (ii) imprint the unknown parameter $h$ onto the probe, (iii) perform a measurement yielding outcomes $\mu_i$, (iv) apply statistical inference (e.g., maximum likelihood) to produce the estimate $\Theta[\mu]$.
  }
  \label{fig:Sensing-protocol}
\end{figure}
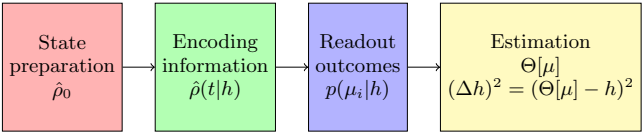

In the standard parameter estimation protocol, one follows four key stages -- as illustrated in Fig.~\ref{fig:Sensing-protocol}:
{\renewcommand\labelitemi{}
\begin{itemize}
    \item (i) prepare the sensor (e.g., an ensemble of atoms) in a known initial state $\hat{\rho}_0$;
    \item (ii) let it interact with the physical system to encode information about the unknown parameter $h$ onto the sensor state, thus transforming for a finite time evolution into $\hat{\rho}_0 \rightarrow \hat{\rho}(t|h)$;
    \item (iii) measure the sensor, yielding an outcome $\mu$ with probability $p(\mu|h)$;
    \item (iv) apply statistical inference (e.g., maximum likelihood estimation) to the collected data to produce a final estimate, $\Theta[\mu]$, of the true parameter $h$, with uncertainty $(\Delta h)^2 = (\Theta[\mu]-h)^2$.
\end{itemize}
}
A natural question arises: what is the best precision achievable in such a protocol? This clearly depends on various factors, such as the initial state preparation, the measured observable, and the employed estimator, among others.
Nevertheless, a very important and powerful result in estimation theory is that, for unbiased estimators (those for which the statistical average yields the true parameter value), the uncertainty is fundamentally limited by the Cram\'{e}r-Rao bound:
\begin{equation}
    \Delta h \ge \frac{1}{ \sqrt{M \mathbb{F}_h(t)}}.
\end{equation}
where $\mathbb{F}_h(t)$ is the Fisher information (FI), and $M$ is the number of independent repetitions of the protocol. The Fisher information, which bounds the maximum precision of the sensor, is defined as follows:
\begin{equation}
    \mathbb{F}_h(t) = \sum_{\mu} \frac{1}{p(\mu|h)} \Bigl(\partial_{h} p(\mu|h)\Bigr)^2.
\end{equation}
 Intuitively, the FI quantifies how sensitive the probability distribution is to small changes in the parameter $h$. A distribution that varies dramatically with this parameter provides a strong signal for estimation, resulting in a higher FI.
 Among all unbiased estimators, the \emph{maximum likelihood estimator} is known to asymptotically saturate the above bound, meaning that in the limit of a large number of independent measurements ($M \gg 1$) its uncertainty approaches $1/\sqrt{M \mathbb{F}_{h}(t)}$.

Within the quantum formalism, the conditional probability distribution originates from measuring a set of Positive Operator-Valued Measure (POVM) operators, $\{\hat M_{\mu}\}$,  which satisfy $\sum_\mu \hat M_\mu = \mathbb{I}$ and
$p(\mu|h) = \Tr(\hat M_\mu \hat \rho(t|h))$. Optimizing over the set of POVM observables, one arrives at the ultimate precision limit for the metrology protocol, defined by the QFI:
\begin{equation}
    F_h(t) \;=\; \max_{\{\hat M_{\mu}\}} \mathbb{F}_h(t).
\end{equation}
The QFI thus represents the maximum amount of information about $h$ that can be extracted from the quantum state, using the most clever measurement observable in the protocol. For a pure state $|\psi_h(t) \rangle$, this quantity takes a simplified form:
\begin{equation}
    F_h(t)
    = 4 \left[\langle \partial_{h} \psi(t) | \partial_{h} \psi(t) \rangle
      - \big|\langle \psi(t) | \partial_{h} \psi(t) \rangle\big|^2\right].
\end{equation}

Alternatively, the QFI can be written by the variance of a ``Heisenberg signal operator'' $\hat{S}$ (HSO), which encodes the entire dynamics of the parameter imprinting:
\begin{equation}
    F_h(t)/4 = \langle \psi(0)|\hat S_h(t)^2 |\psi(0) \rangle - \langle \psi(0)| \hat S_h(t) |\psi(0) \rangle^2
\end{equation}
This operator is defined as the time-integrated derivative of the sensor's Hamiltonian, expressed in the Heisenberg picture,
\begin{equation}
\label{eq.HSO.definition}
    \hat S_h(t) = \int_0^t \hat U^\dagger(t') \left( \frac{\partial \hat H(t')}{\partial h}  \right) \hat U(t') \, dt'
\end{equation}
with $\hat H(t)$ the sensor Hamiltonian and $\hat U(t) = \mathcal{T}e^{-i\int_0^t \hat H(t') dt'}$ is the corresponding unitary evolution operator.

The scaling of the Fisher Information with time and system size highlights a fundamental advantage of quantum and/or correlated sensors. In classical systems, information accumulates linearly with the duration of the experiment ($F_h \propto t$). In contrast, the coherent nature of quantum evolution can lead to a quadratic scaling ($F_h \propto t^2$) for a single quantum particle. When employing an ensemble of $N$ uncorrelated particles, and the signal acting independently on each of them, the information scales as $F_h \propto Nt^2$. This scaling can be enhanced using entangled states, where particle correlations can lead to a scaling proportional to the square of the number of particles, $F_h \propto N^2t^2$, representing the ultimate limit for quantum-enhanced metrology, also known as the Heisenberg limit.

\section{Sensor Hamiltonian}

In this section, we introduce the specific sensor Hamiltonian and detail its key properties. We consider a quantum sensor consisting of a one-dimensional chain of $N$ spin-$1/2$ particles, featuring the MBS phase. Then the sensor is coupled to an external periodic field, and our objective is to estimate the unknown amplitude $h$ of this field with high precision. The total Hamiltonian governing the system is
\begin{equation}
\hat{H}(t) = \hat{H}_{\rm scar} + \hat V(t),
\end{equation}
where $\hat{H}_{\rm scar}$ is the internal, time-independent Hamiltonian responsible for generating the many-body scar dynamics. The coupling to the external field is described by $\hat V(t) = h \sin{(\omega t)} \hat O$, where $h$ is the target amplitude, $\omega$ is a known frequency, and the operator $\hat O$ defines the structure of the coupling to the sensor.

Our work focuses on many-body scars generated by the paradigmatic PXP Hamiltonian \cite{turner2018weak}, given by
\begin{equation}
\hat H_{\rm scar} \equiv \hat H_{PXP} = \sum_{i=1}^N \hat{P}_{i-1} \hat \sigma_i^x \hat P_{i+1}.
\end{equation}
Here, $ \hat \sigma_i^\alpha$ (with $\alpha=x,y,z$) are the Pauli matrices for the $i$-th spin, and $ \hat P_i = |\downarrow \rangle \langle \downarrow |_i$ is the projector onto the local spin-down state.
 We consider a chain with periodic boundary conditions. Furthermore, we will consider two types of external field operator: (i) total spin magnetization along the z-direction, $\hat O \equiv \hat{O}^z =\sum_i \hat \sigma_i^z$, or (ii)  staggered magnetization $\hat O \equiv \hat{O}^{stag} = \sum_i(-1)^{i}\hat \sigma_i^z$.

It is pertinent to note that while our specific analysis centers on the PXP model, a compelling future direction is to explore whether other systems or mechanisms for scarring can likewise be harnessed for quantum-enhanced sensing. Here, the PXP model possesses both an \textit{approximate} spectrum-generating algebra—namely, a set of operators that approximately connect eigenstates in a ladder-like structure with nearly uniform energy spacing—and a Krylov subspace structure \cite{saad1989overview,kressner2010krylov} which represent two general frameworks for understanding scarred dynamics. We expect that, at least at a qualitative level, the enhanced sensing properties we report extend to systems where scars arise from these mechanisms.

The many-body scars in the PXP model have a few peculiar properties that are directly relevant to the sensor performance. Precisely, given a specific initial state preparation in a period-$2$ charge density wave,
\begin{equation}
|Z_2 \rangle = |\uparrow \downarrow ... \uparrow \downarrow  \rangle,
\end{equation}
contrary to the expectations of the eigenstate thermalization hypothesis (ETH)—which states that, for generic interacting systems, individual energy eigenstates behave thermally and local observables relax to their equilibrium values \cite{deutsch1991quantum, srednicki1994chaos, rigol2008thermalization}, this state does not thermalize rapidly. Instead, its dynamics are characterized by a long-lived spin coherence followed by periodic revivals \cite{turner2018weak}. 

\begin{figure}
    \centering
    \includegraphics[width=1\linewidth]{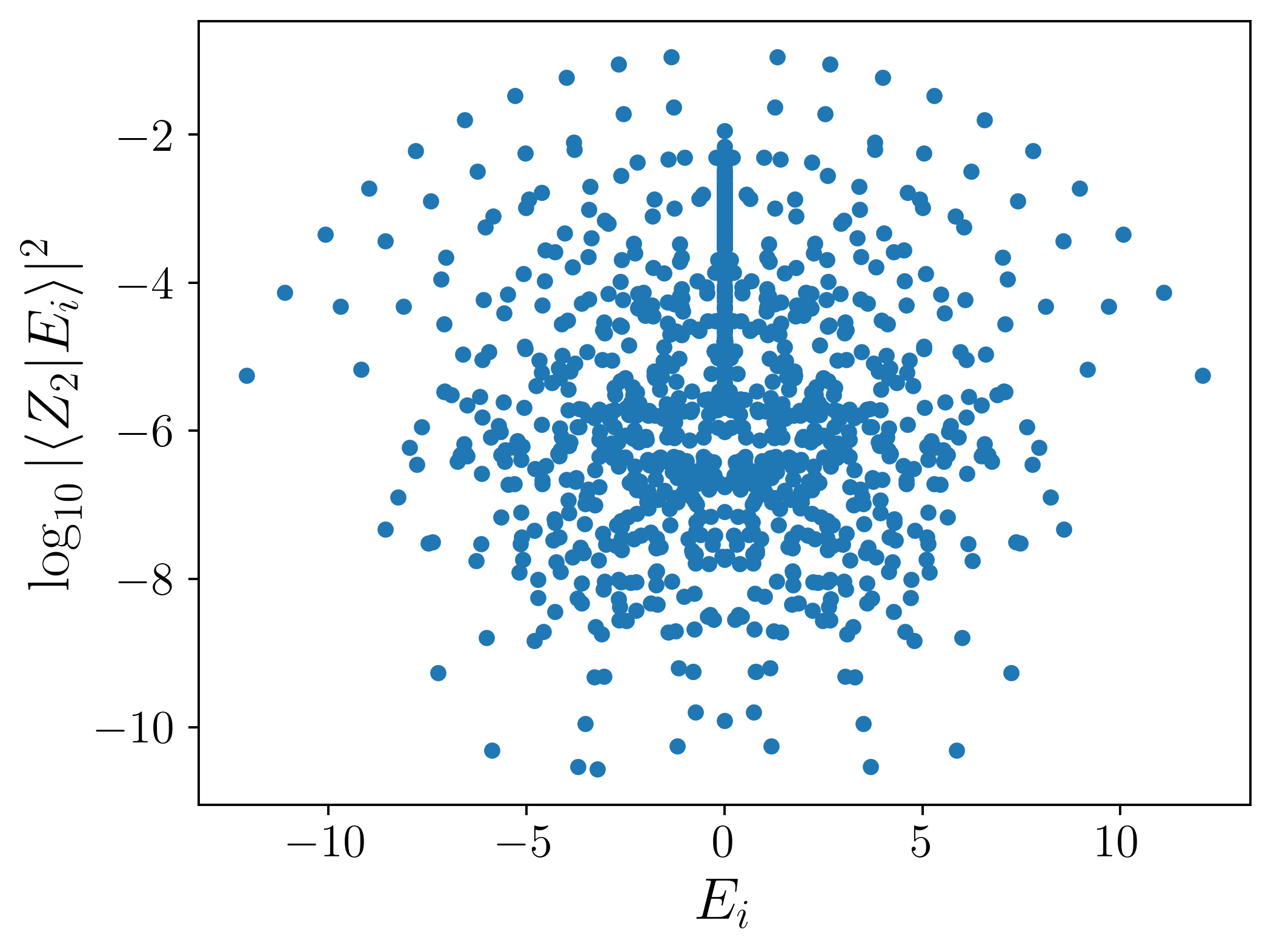}
    \caption{\textit{MBS tower.- } Overlap of the PXP eigenstates $\{|E_i\rangle\}$ with the $|Z_2\rangle$ state for a chain with $N=20$ spins. The tower structure, with its distinct layers, is clearly visible, with the highest-overlapping eigenstates spaced roughly equally in energy around the center of the spectrum.
    }
    \label{fig:spectrum}
\end{figure}

This weak ergodicity breaking arises from the presence of a special set of non-thermal eigenstates in the many-body spectrum that have an atypically high overlap with the $|Z_2 \rangle$ initial state. 
 The structure can be clearly observed by decomposing the $|Z_2 \rangle$ state over the eigenbasis $\{|E_i\rangle\}$ of the PXP Hamiltonian, as shown in Fig.~\ref{fig:spectrum}. The decomposition reveals a distinctive "tower" of scar states spread across the spectrum. These eigenstates possess a sub-extensive entanglement entropy, distinguishing them from their thermal counterparts. Notably, the energy separation between consecutive scar states near the center of the spectrum tends to an approximately constant value by increasing the system size, with an energy gap of $(\Delta E)\sim 1.337$. Moreover, this tower structure manifests across multiple energy layers. 
 While finite-size effects are present, both the energy gap between scar states within a given layer converges to $(\Delta E)$, and the number of visible tower layers increases as the system size grows.

It is worth remarking that the dynamics in the PXP model are highly constrained. A spin flip at a given site is permitted only if both neighboring sites are projected into the down state. This kinetic constraint means that, for specific initial preparations like $|Z_2 \rangle$, vast portions of the full Hilbert space are dynamically inaccessible. In fact, the evolution starting from $|Z_2 \rangle$ is strictly confined to a subspace known as the Fibonacci subspace, whose dimension scales as $\mathcal{D}_N = f_{N-1} + f_{N+1}$, where $f_N$ is the $N$'th Fibonacci number and $N$ is the length of the chain. Consequently, the effective Hilbert space dimension scales as $\mathcal{D}_N \sim \phi^N$, where $\phi \approx 1.618$ is the golden ratio. Although it remains exponentially large with $N$, this effective dimension is substantially smaller than the  full Hilbert space dimension of $2^N$, thus amenable for exact numerical calculations of chains with $N \approx 20$ spins \cite{turner2018weak}.

\section{Sensor performance}
In this section, we present our analysis of the scar sensor's performance in AC fields.
We explore its maximum precision via the QFI for different AC field structures, showing how to tune it to achieve quantum-enhanced performance. Furthermore, we provide a semi-analytical expression for the QFI, allowing us to gain further insights and extrapolate our results to macroscopic sensors.

\subsection{Exploring coherent time in weak-thermalizing sensors}

We begin our analysis with a critical examination of sensor resources and how to exploit them optimally in realistic quantum metrology protocols. As previously mentioned, the fundamental Heisenberg limit, characterized by QFI scaling as $F_h(t) \propto N^2t^2$, represents the ultimate sensitivity bound. Achieving this requires a sensor that maximally leverages both extensive entanglement across $N$ spins and prolonged interrogation time $t$. However, the highly entangled states required to boost the $N$-scaling, such as GHZ states, are generally fragile and susceptible to decoherence \cite{huelga1997improvement, escher2011general}. Consequently, such a sensor can only sustain the Heisenberg scaling for a finite coherence time $t_*$, after which its performance deteriorates into a sub-Heisenberg regime (i.e., a scaling smaller than quadratic in $N$ and/or $t$).

Assuming that the performance of such a Heisenberg-limited sensor is severely deteriorated after $t_*$ (leading to a negligible sub-Heisenberg scaling), a practical strategy is to operate it only within its coherent lifetime. Specifically, by repeatedly initializing the protocol and measuring over $M = T/t_*$ independent cycles with a total duration of $T$, one can fully explore its optimal scaling. The resulting Cram\'{e}r-Rao bound on the estimation error for the parameter $h$ is then given by
$(\Delta h)_{\rm Heis,t} = \sqrt{ 1 / ( M N^2 t_*^2 ) }$; using
$M = T/t_*$, the expression becomes
$(\Delta h)_{\rm Heis,t} = \sqrt{ 1 / ( N^2 t_* T ) }$.

Conversely, sensors that employ less correlated states and exhibit a sub-Heisenberg scaling of the form $F_h(t) = N^\alpha t^\beta$ (with $\alpha<2$ and/or $\beta<2$) may prove advantageous if they possess significantly extended coherence lifetimes, a feature, for example, associated with non-equilibrium phases exhibiting weak ergodicity breaking. For such a sensor operating in a single continuous measurement of duration $T$ ($M=1$), the estimation uncertainty is $(\Delta h)_{\rm ss} = 1 / \sqrt{ N^\alpha T^\beta }$. The ratio of the uncertainties,
\begin{equation}
\frac{(\Delta h)_{\rm Heis,t_*}}{(\Delta h)_{\rm ss}} = \sqrt{ \frac{ N^{\alpha-2} T^{\beta-1} }{ t_* } },
\end{equation}
reveals that the single-shot sensor outperforms the repeated Heisenberg strategy when \begin{equation}
T^{\beta-1} > t_* N^{2-\alpha}.
\label{metrological-regime}
\end{equation}
This condition highlights that when decoherence severely limits the interrogation time of a Heisenberg-limited sensor, probes based on weakly thermalizing dynamics may provide a competitive or even superior metrological performance. In particular, experiments have shown many-body scar systems to exhibit persistent revivals and slow thermalization for specific initial states, indicating the presence of extended coherent dynamics beyond generic many-body timescales \cite{turner2018weak}. This behavior contrasts with highly entangled probes such as GHZ states: although GHZ states can, in principle, achieve Heisenberg-limited sensitivity, their optimal interrogation time is strongly constrained by realistic local decoherence, leading to a rapid deterioration of their metrological advantage as system size increases \cite{demkowicz2012elusive}.

Substituting this behavior into the above condition highlights that the relative performance
of the two strategies is controlled by the scaling of the optimal interrogation time with system
size. In particular, when decoherence causes the optimal interrogation time of a Heisenberg-limited
sensor to shrink with increasing $N$, the advantage of quadratic scaling in $N$ can be effectively
suppressed. In this regime, a weakly thermalizing sensor with correlation-enhanced scaling
($\alpha>1$) and extended coherent dynamics can outperform fragile Heisenberg-limited probes
in the many-body limit ($N\gg1$).

To make this condition more explicit, let us consider a weakly thermalizing sensor characterized by a scaling $\alpha = 1+\epsilon$, $\epsilon \ll 1$ reflecting the presence of weak many-body correlations, and $\beta=2$  with an extended coherence. The QFI thus scales as  $F_h \sim N^{1+\epsilon} t^2$. In this case, the condition of Eq.~\eqref{metrological-regime} reduces to
\begin{equation}
T > t_* N^{1-\epsilon}.
\end{equation}
For Heisenberg-limited probes such as GHZ states, the optimal interrogation time typically decreases with system size as
$t_* \sim N^{-\gamma}$, with $\gamma \gtrsim 1$. Substituting this behavior yields
$T > N^{1-\epsilon-\gamma}$, which is readily satisfied in the many-body limit.
This demonstrates that even weakly correlated sensors ($\epsilon \ll 1$), provided they exhibit extended coherent dynamics, could outperform Heisenberg-limited strategies.
This observation indicates that robustness against thermalization, rather than maximal entanglement, may constitute the more relevant resource for scalable quantum metrology.

\subsection{Heisenberg signal operator}

We are interested in sensor operation for detecting extremely weak AC fields, in which the signal amplitude $h$ is negligible compared with all other system parameters.  In this linear response regime ($h \rightarrow 0$), the QFI
can be computed more easily via its connection to the Heisenberg signal operator (Eq.~\eqref{eq.HSO.definition}). The expansion of the HSO in the eigenbasis of the scar Hamiltonian $\hat{H}_{\rm scar}$ reduces it to the simpler form~\cite{Tsypilnikov2026,PhysRevA.109.L050203}:
\begin{equation}
    \hat{S}(t) = \sum_{i,j} O_{ij} R_{ij}(\omega,t) \ket{E_i}\bra{E_j}
    \label{eq_fisher_final}
\end{equation}
Here, $O_{ij} = \langle E_i| \hat{O} | E_j\rangle$ represents the matrix element connectivity of the AC field operator $\hat{O}$ across the many-body spectrum. The time-dependent response amplitude is governed by the function
\begin{equation}
 R_{ij}(\omega,t) = \left(
    \frac{e^{i(\Delta E_{ij}-\omega )t}-1}{(\Delta E_{ij}-\omega)}-\frac{e^{i(\Delta E_{ij}+\omega )t}-1}{(\Delta E_{ij}+\omega )}\right)/2
    \label{eq_R_ij}
\end{equation}
where $\Delta E_{ij} = E_i - E_j$ is the energy gap between eigenstates.

The ultimate sensitivity, bounded by the norm of $\hat{S}(t)$, is thus contingent upon maximizing both the connectivity and response terms. Moreover, recalling that the QFI is the variance of the HSO evaluated in the initial state, its value is dominated by contributions from eigenstates with the largest overlap with that state; in our case, these are those within the MBS tower. We therefore analyze the connectivity and response terms with a specific focus on their properties along the tower.

To be clearer, we will use a notation that explicitly discriminates the eigenvalues along the different layers of the MBS tower. Specifically, from now on, we will use the double-index notation:
\begin{equation}
    |E_i\rangle \rightarrow |E_{j}^{\ell}\rangle
\end{equation}
with the superscript index $\ell=1,2,...$ denoting the layer of the tower, ordered from the most dominant one (highest overlap with the initial state, $\ell=1$) to those with lower overlaps, and $j=0,1,-1,2,-2,...$ denotes its elements. We define $j=0$ corresponding to the eigenvalue at the center of the layer, $E_{j=0}^{\ell} = 0$, and those with $j$ negative (positive) are the ones to the left (right) of it. 
Likewise, the energy gaps are denoted by,
\begin{equation}
\Delta E_{ij} \rightarrow \Delta E_{j,j'}^{\ell,\ell'} = E_{j}^{\ell} -  E_{j'}^{\ell'} 
\end{equation}
using the same double-index convention as above. Many times, however, we will be only interested in the properties of the first layer. In these cases, for simplicity, we omit the double index notation and simply use $E_i^1 \rightarrow E_i$. 

\textit{Response terms.-} It is direct to observe that the response function $R_{ij}(\omega,t)$ for a single pair of states is maximized close to resonance, $\omega \sim \Delta E_{jj'}^{\ell \ell'}$, where it grows linearly with time till an off-resonance time scale 
$t_{\rm off}(\omega) = (\omega -  \Delta E_{jj'}^{\ell \ell'})^{-1}$. This is not surprising, since a resonant field is known to induce a coherent tunneling transition among its corresponding pair of states. The key difference
of the scar-based sensor, however, lies in the structure of the tower: by tuning the AC field frequency to a multiple of the characteristic energy spacing $\omega = m (\Delta E)$, $m\in \mathbb{Z}$, one can now simultaneously activate a resonant response across a macroscopic number of state pairs constituting the layers of the tower.
This collective, many-body resonance can bring new features and potentially amplify the overall response amplitude.

\begin{figure*}
\includegraphics[width=0.45\linewidth]{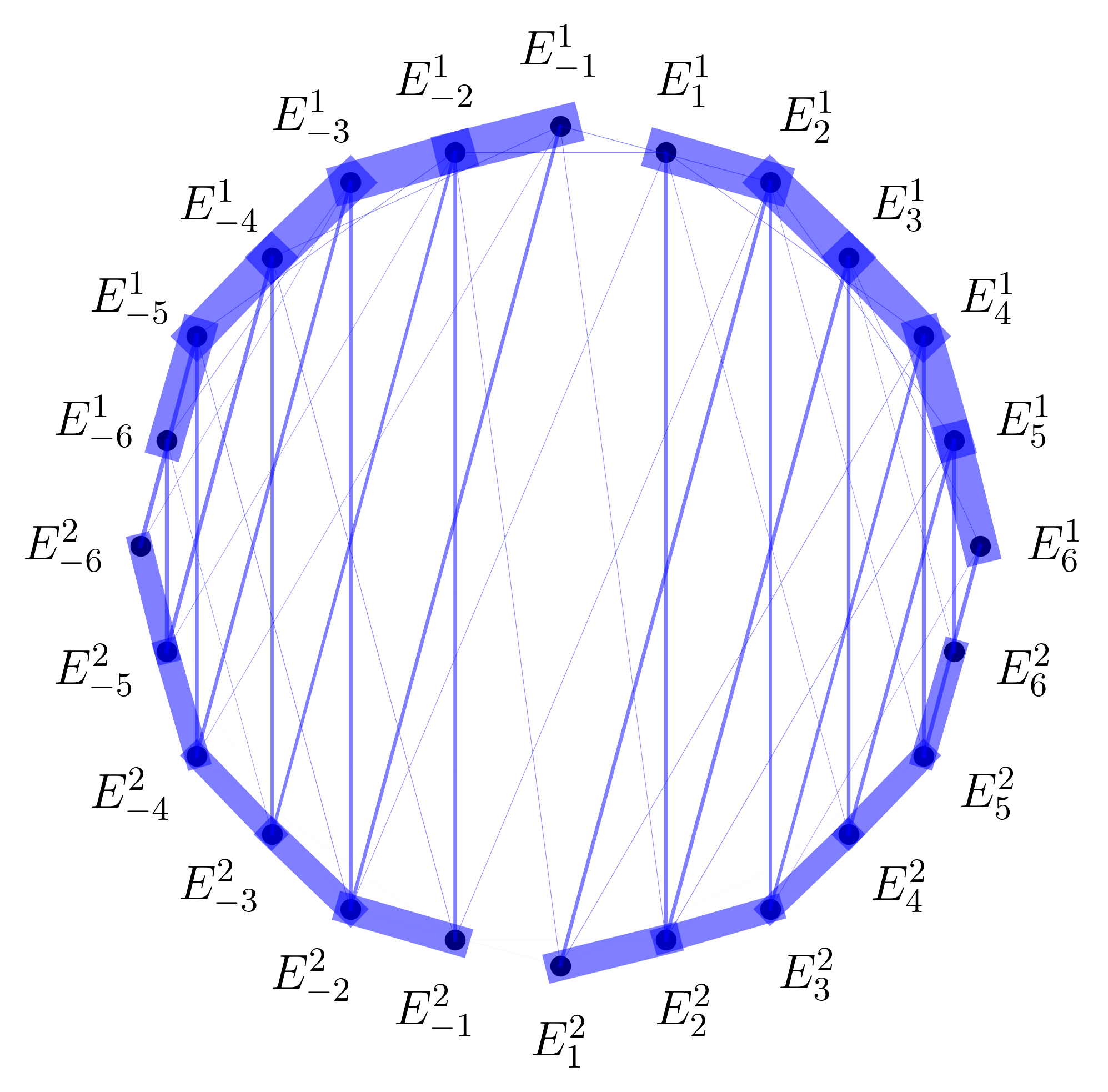}\includegraphics[width=0.45\linewidth]{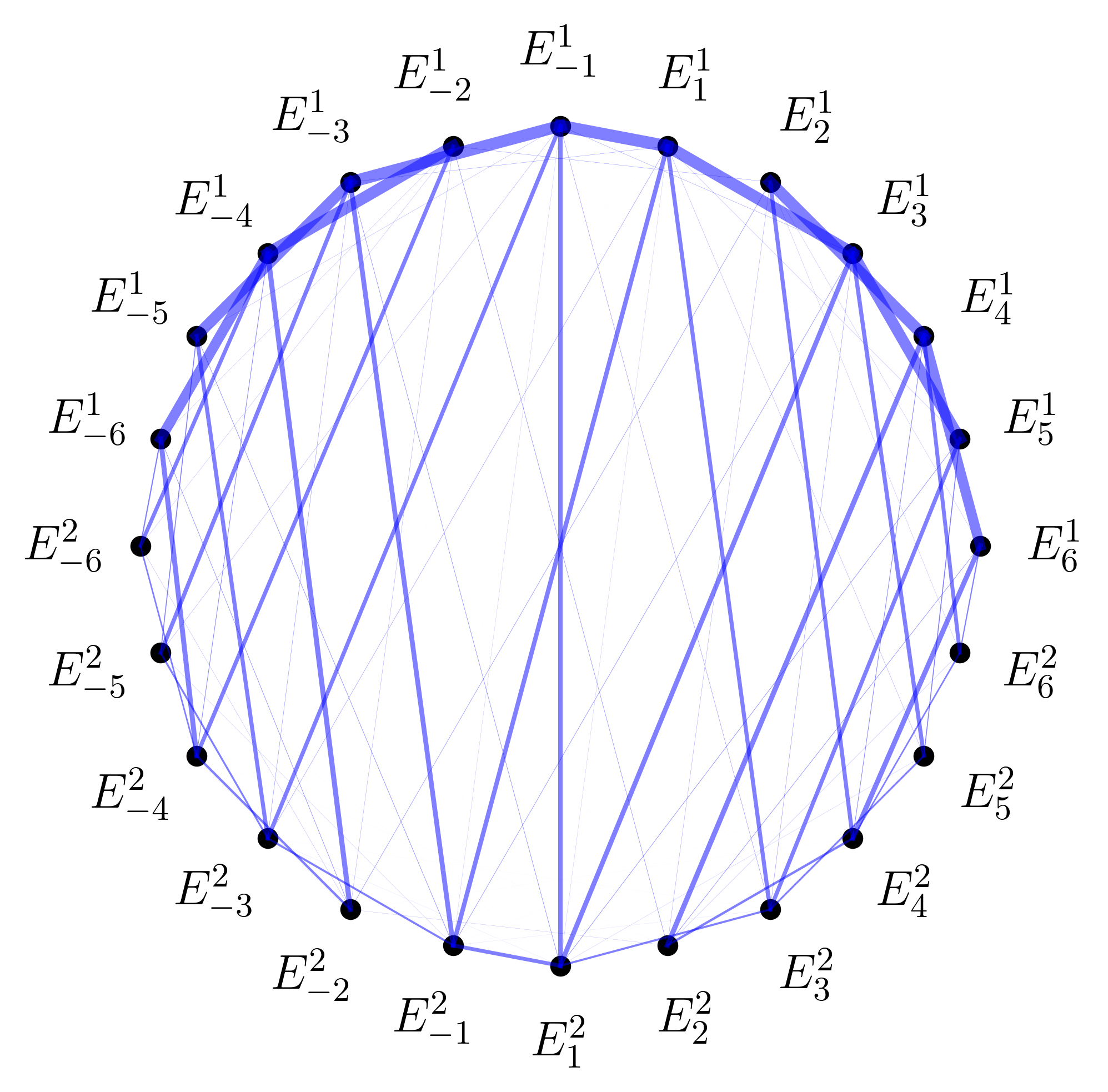}
\caption{The graph shows the connectivity $(\hat O)_{jj'}^{\ell \ell'}$ between the different scar states with $N=20$ spins, considering either \textbf{(a)} a staggered field ($\hat O^{\rm stag}$), or \textbf{(b)} a homogeneous magnetization ($\hat O^z$). Thicker lines in the graph correspond to larger absolute values. Here, we show the connectivity among the first two main layers in the MBS tower spectrum.
We see that the connections are significantly stronger among nearest-neighbor (next-nearest-neighbor) eigenvalues for the staggered (homogeneous) magnetization operator, both within and between layers.
}
\label{fig:connection_terms}
\end{figure*}

\textit{Connectivity.- } We find that the spatial profile of the field operator $\hat{O}$ dictates the connectivity pattern, as illustrated in
Figs.(\ref{fig:connection_terms}a)-(\ref{fig:connection_terms}b). Specifically, a staggered magnetization perturbation predominantly couples nearest-neighbor eigenstates along the layers of the tower. These couplings are most dominant along the same layer, but also with non-negligible connections between different layers. In contrast, a homogeneous magnetization operator features dominant matrix elements connecting next-nearest-neighbor states.

\begin{figure}
\includegraphics[width=0.5\linewidth]{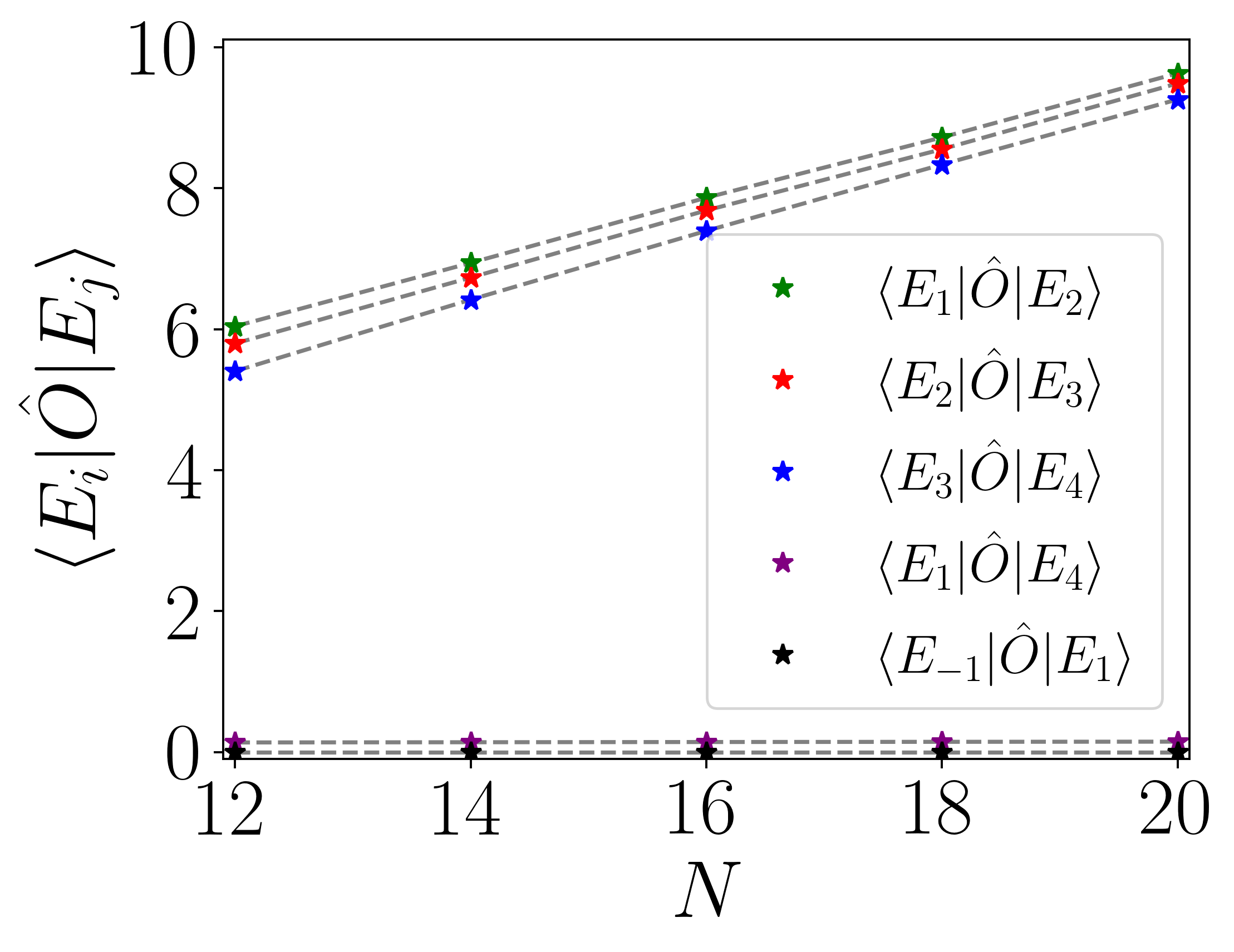}\includegraphics[width=0.5\linewidth]{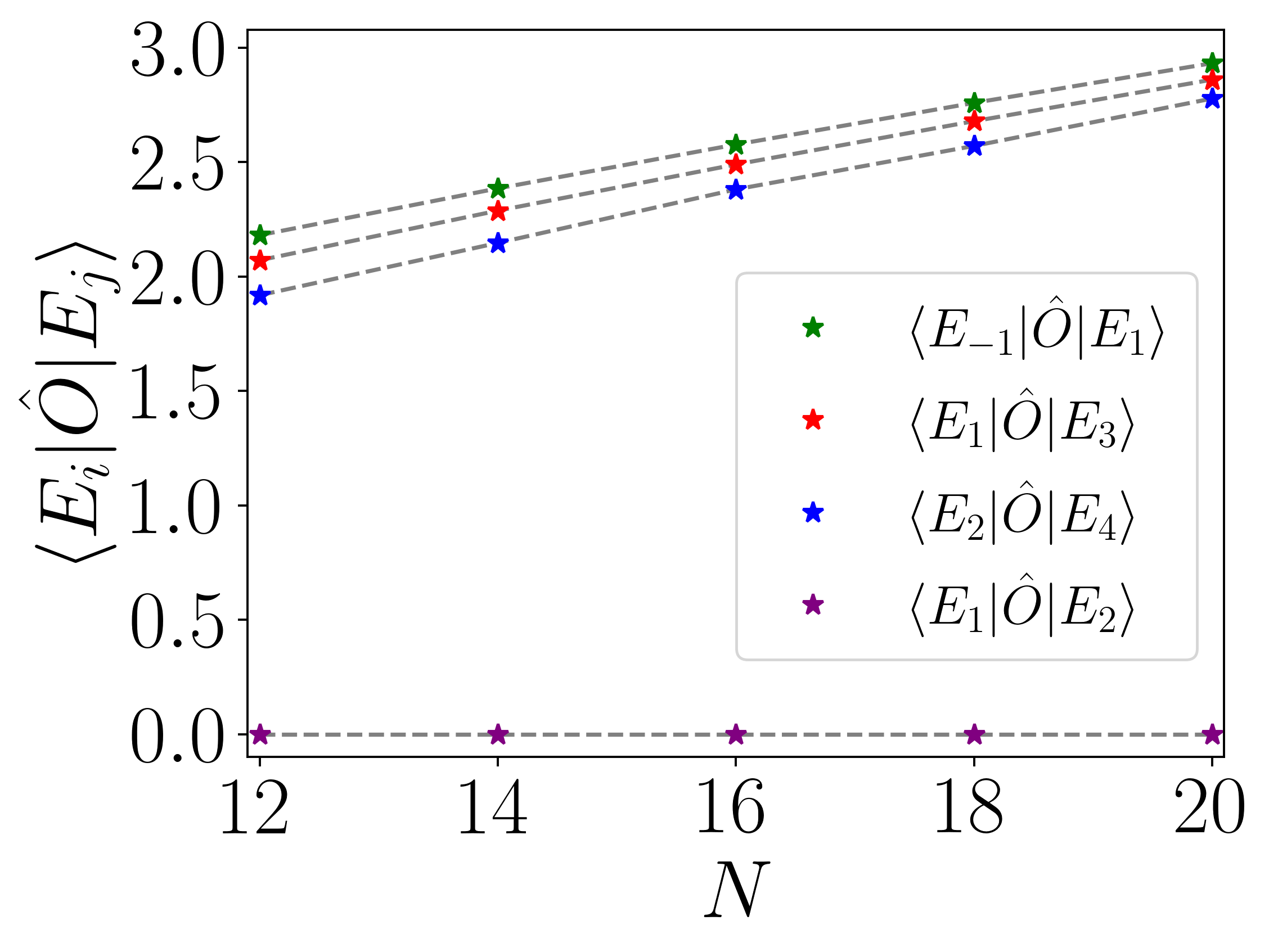}
\caption{Finite-size scaling analysis for the connectivities along the first layer, considering \textbf{(a)} a staggered field ($\hat O^{\rm stag}$), or \textbf{(b)} a homogeneous magnetization ($\hat O^z$). While the connectivity among nn (nnn) eigenvalues for a staggered (homogeneous) magnetization grows polynomially with the system size, others are negligible and do not scale significantly with $N$.}
\label{fig:connection_terms_fss}
\end{figure}

Moreover, while the connectivity across the main connections (nn or nnn) scales polynomially with system size \( N \), others remain negligible and do not scale considerably with the system size - as shown in Figs.~\ref{fig:connection_terms_fss}(a)-(b) - underscoring their relevance for the sensor analysis. We extract explicit scaling forms for these dominant matrix elements. For the staggered magnetization operator, the nearest-neighbor coupling along the first layer ($\ell=1$) of the tower scales as
\begin{equation}
(O^{\mathrm{stg}})_{i,i+1}^{1,1} = k_{\mathrm{stg}} N^{\gamma_{\mathrm{stg}}},
\label{eq:scaling_stg}
\end{equation}
with \( k_{\mathrm{stg}} \approx 0.62 \) and \( \gamma_{\mathrm{stg}} \approx 0.91 \). Conversely, for the total (homogeneous) magnetization, the dominant next-nearest-neighbor matrix element follows
\begin{equation}
(O^{\mathrm{tot}})_{i,i+2}^{1,1} = k_{\mathrm{tot}} N^{\gamma_{\mathrm{tot}}},
\label{eq:scaling_tot}
\end{equation}
where \( k_{\mathrm{tot}} \approx 0.52 \) and \( \gamma_{\mathrm{tot}} \approx 0.58 \). It is worth noting the significantly larger exponent in the staggered case, reflecting that the operator’s spatial profile selectively enhances specific resonant pathways, thereby governing the different dynamical responses of the tower of states.

\subsection{Frequency Scanning}

\begin{figure}
    \centering
        \begin{overpic}[width=1\linewidth]{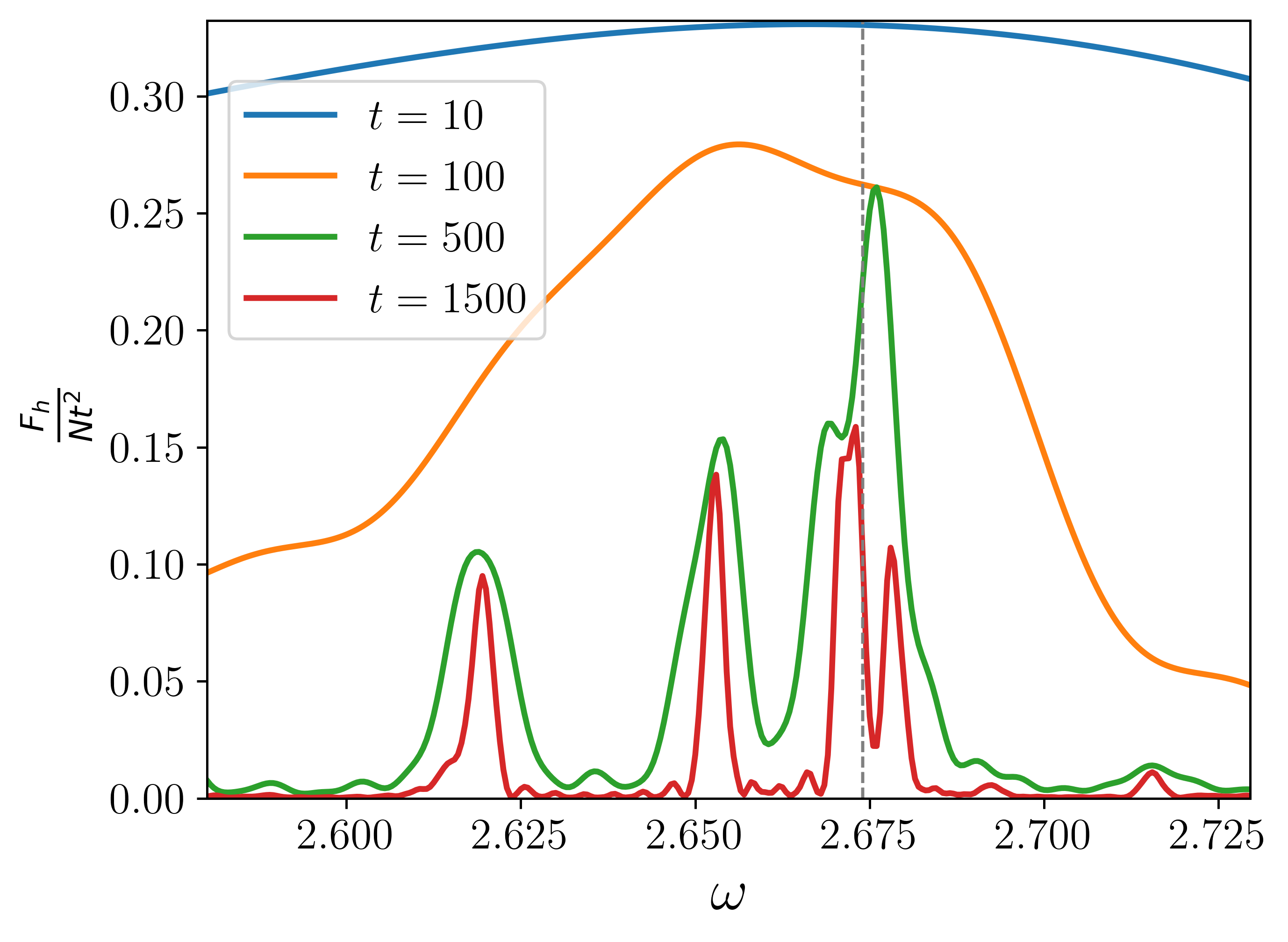}
        \put(0,60){\bfseries\large (a)}
    \end{overpic}
        \begin{overpic}[width=1\linewidth]{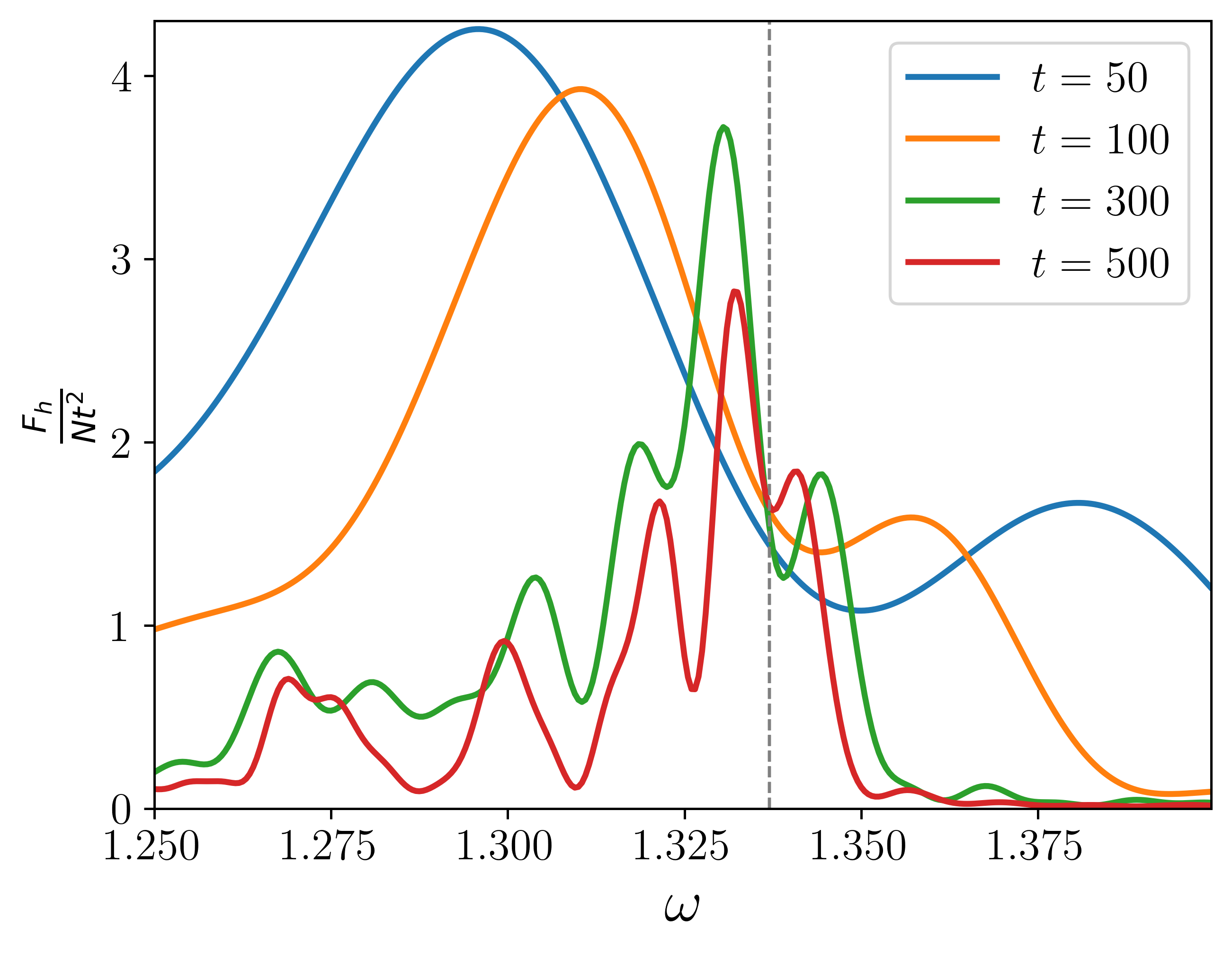}
        \put(0,60){\bfseries\large (b)}
    \end{overpic}
    \caption{Scanning frequency of the QFI around its resonance values, considering 
    either \textbf{(a)} a homogeneous magnetization ($\hat O^z$), or \textbf{(b)} a staggered field ($\hat O^{\rm stag}$). The results are shown for a system with $N=20$ spins.
    }
    \label{fig:scanning_vartime}
\end{figure}

The sensor's efficacy depends critically on the relationship between the AC frequency and the tower's resonant gaps. Based on our previous discussion on response terms and connectivity, the potentially most suitable AC frequencies lie on the nn-resonance $\omega_{\rm stg} = \Delta E$ for staggered magnetization, and nnn-resonance $\omega_{\rm tot} = 2 \Delta E$  for homogeneous one. Therefore, we perform a frequency-scanning analysis of the QFI dynamics around these values. Our results are shown in Fig.~\ref{fig:scanning_vartime}.

We first observe that the peak of the QFI tends to converge towards the system's resonance frequency ($\omega_{\rm stg}$ or $\omega_{\rm tot}$) over time. Consequently, to exploit sensors with long interrogation times, the driving frequency should be tuned into resonance with the tower's spectral structure. Operating on-resonance ensures the most efficient amplification of the target signal over extended periods. Worth remarking that not only a dominant peak at the resonance frequency emerges for long times ($t \approx 500$ in the figures), but also a few more smaller ones around it. This is because the gaps between nn or nnn eigenenergies in the layers of the MBS tower are not all exactly equal to the resonance frequency. Due to finite-size effects, the corrections are small, giving rise to multiple smaller peaks in the sensor response. In the macroscopic limit, one thus expects all of these to converge to a single resonant frequency.

Interestingly, we also observe that the off-resonant response can achieve substantial sensitivity. This regime may become relevant in the presence of experimental decoherence or dissipation that limits the apparatus's effective coherence time. If the system's lifetime is shorter than the time required to achieve resonant amplification, operating at a slightly off-resonant frequency (where a significant QFI peak emerges on a shorter timescale) could provide an advantageous trade-off.

\subsection{Finite-Size Scaling}

\begin{figure}
     \centering
             \begin{overpic}[width=1\linewidth]{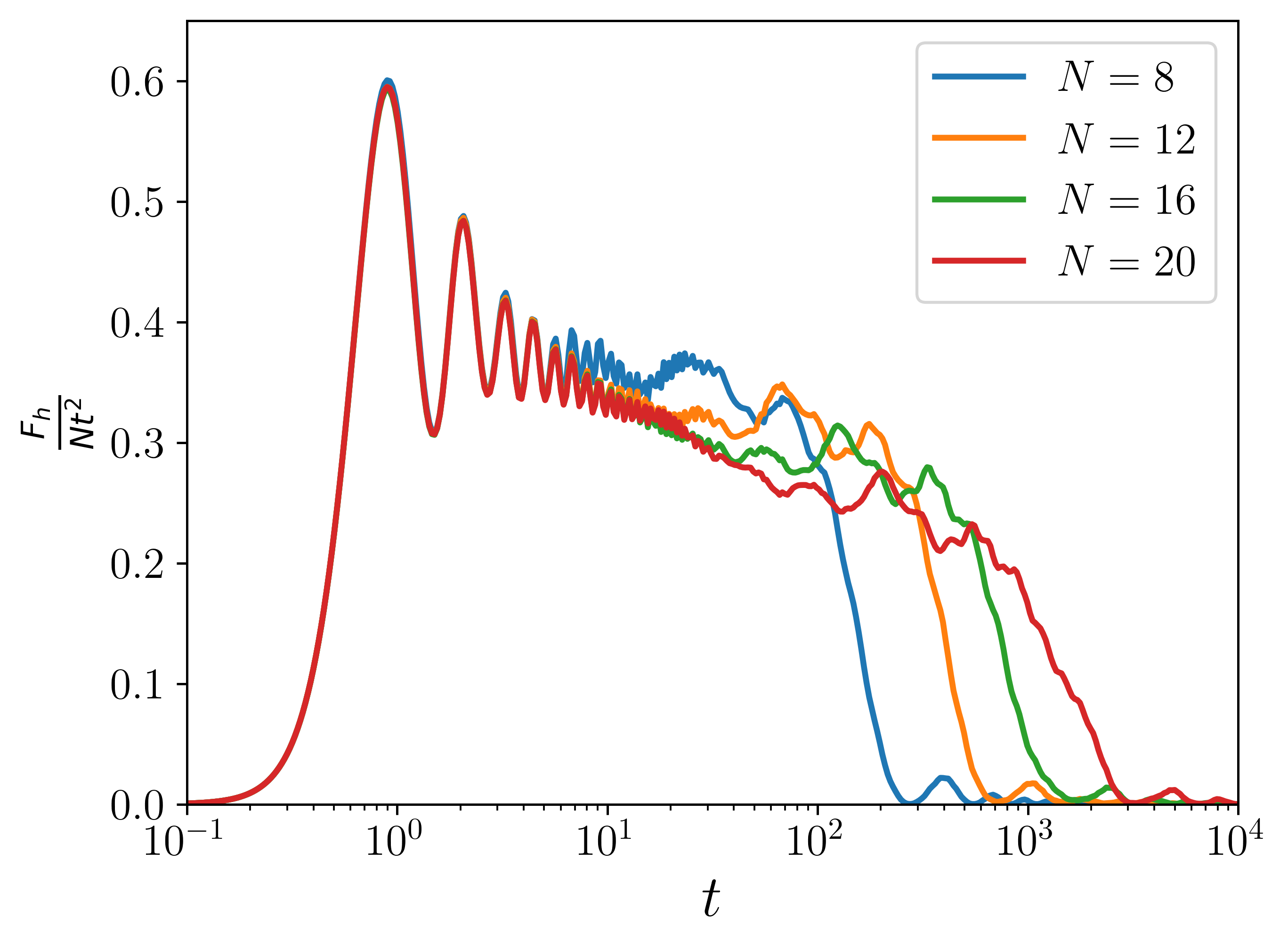}
        \put(0,60){\bfseries\large (a)}
    \end{overpic}

        \begin{overpic}[width=1\linewidth]{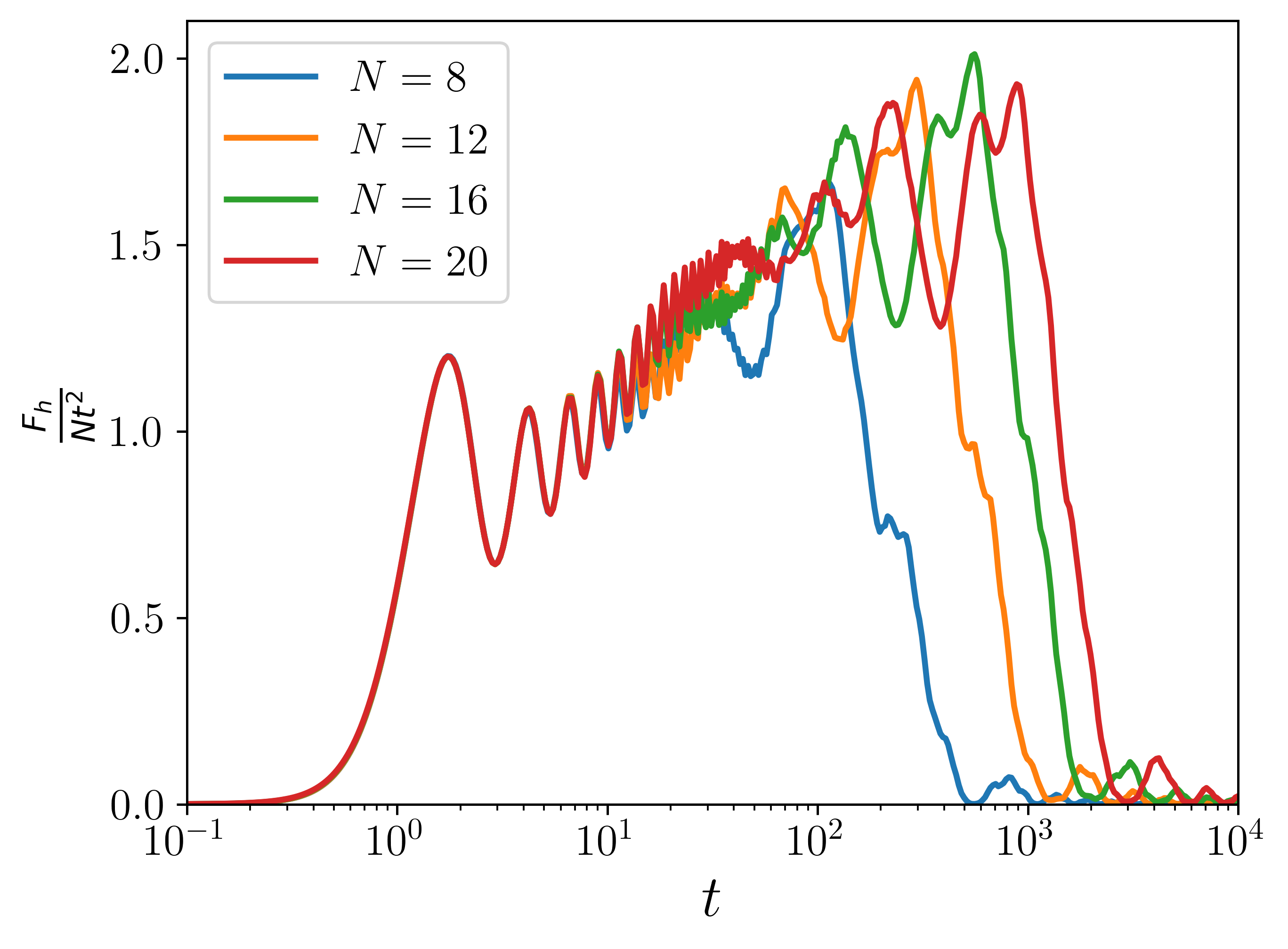}
        \put(0,60){\bfseries\large (b)}
    \end{overpic}
     \caption{ Dynamics of the QFI at resonance frequency, for different system sizes. In \textbf{(a)} we consider the case of a homogeneous magnetization with $\omega = 2 \Delta E$, while in \textbf{(b)} the  staggered field with $\omega = \Delta E$.}
     \label{fig:overtime_resonance}
 \end{figure}

The scaling of performance with system size \(N\) further delineates the different operational regimes. When the sensor is tuned precisely to resonance, the operational time window with the QFI scaling at least as favorable as the standard quantum limit (SQL) is found to increase with the number of spins - as shown in Fig.~\ref{fig:overtime_resonance}. This positive scaling underscores the relevance and potential enhancement of such sensor designs in macroscopic systems, where large \(N\) can be leveraged to extend the coherent sensing period.

\begin{figure}
     \centering
        \begin{overpic}[width=1\linewidth]{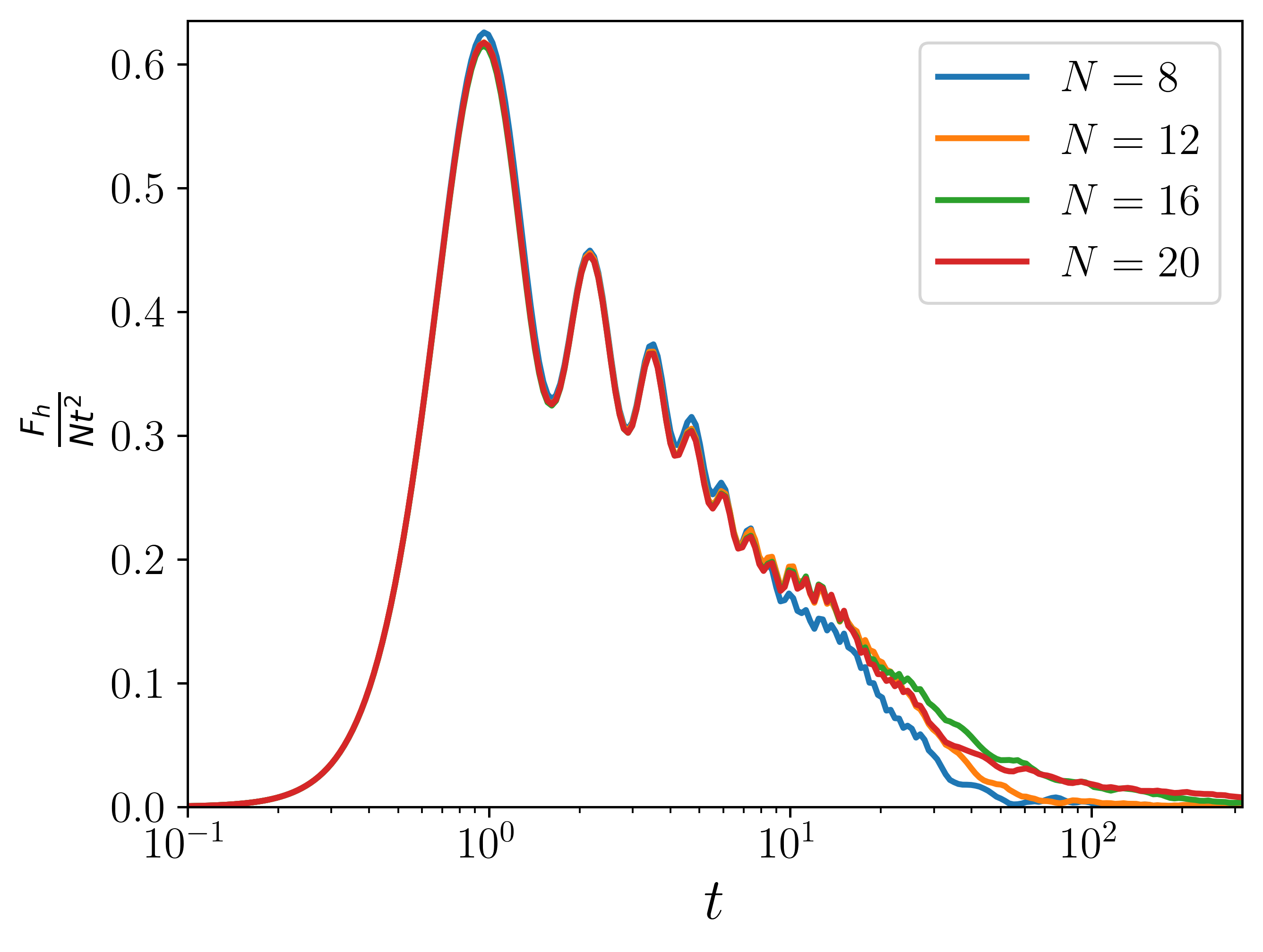}
        \put(0,60){\bfseries\large (a)}
    \end{overpic}

        \begin{overpic}[width=1\linewidth]{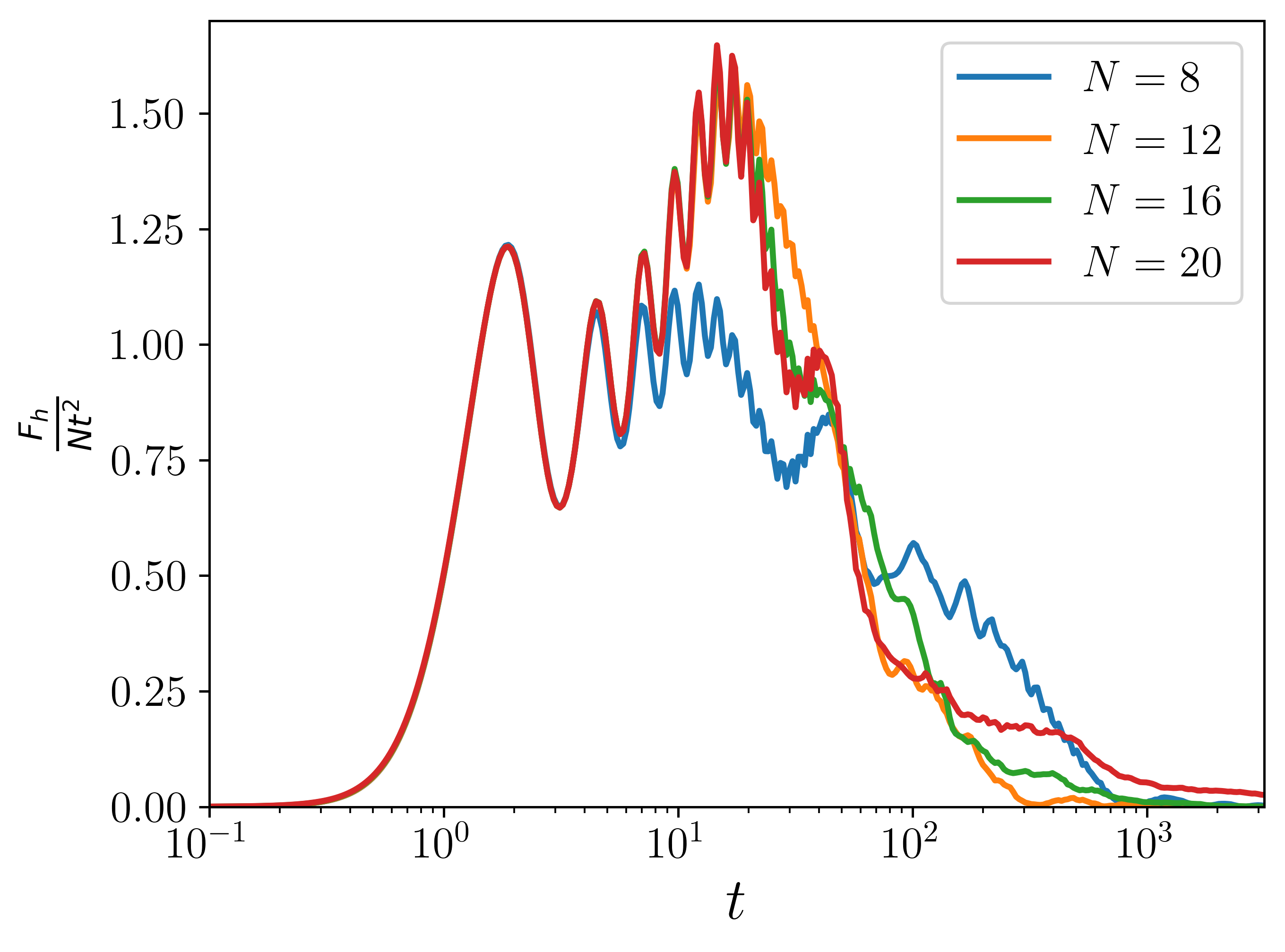}
        \put(0,60){\bfseries\large (b)}
    \end{overpic}
     \caption{ Dynamics of the QFI off-resonantly, for different system sizes. In \textbf{(a)} we consider the case of a homogeneous magnetization with $\omega = 2.2 \Delta E$, while in \textbf{(b)} the staggered field with $\omega = 1.1 \Delta E$. 
     }
     \label{fig:overtime_outRessonance}
 \end{figure}

Conversely, for frequencies slightly off-resonance, the behavior changes markedly, see Fig.~\ref{fig:overtime_outRessonance}. A prominent peak in the QFI emerges, which, for a fixed, comparable timescale, can be significantly larger than the on-resonance value. However, upon increasing the number of spins, the magnitude of this off-resonant QFI peak does not scale favorably with \(N\). This indicates that while off-resonant operation can offer high sensitivity for small systems within short time windows, it may not constitute the optimal strategy in the macroscopic limit.

\section{Single tower approximation}

To gain analytical insight into the dynamics of the QFI and its scalings, we develop an approximation under the resonant condition $\omega = \omega_{\text{stg (tot)}}$, corresponding to the staggered or total magnetization signal. Our approach relies on three key simplifying assumptions that capture the sensor's dominant behavior.

\textit{(i) Operator truncation.} We truncate the Heisenberg signal operator (HSO) to retain only its most dominant matrix elements along the connectivity terms. Concretely, for the staggered (total) magnetization, we restrict the dynamics to nearest-neighbor (next-nearest-neighbor) couplings between eigenstates of the highest tower, effectively projecting onto the subspace defined by these dominant transitions.

\textit{(ii) Uniform level spacing.} We assume the energy spacing between these dominant eigenstates is exactly equal to the many-body scar gap $\Delta E$ for total magnetization (or $2\Delta E$ for staggered one). This neglects finite-size corrections to the level spacings and is formally valid either in the macroscopic limit ($N \to \infty$) or for times short enough that small deviations in the spacings are not resolved.

\begin{figure}
\centering
\begin{overpic}[width=\linewidth]{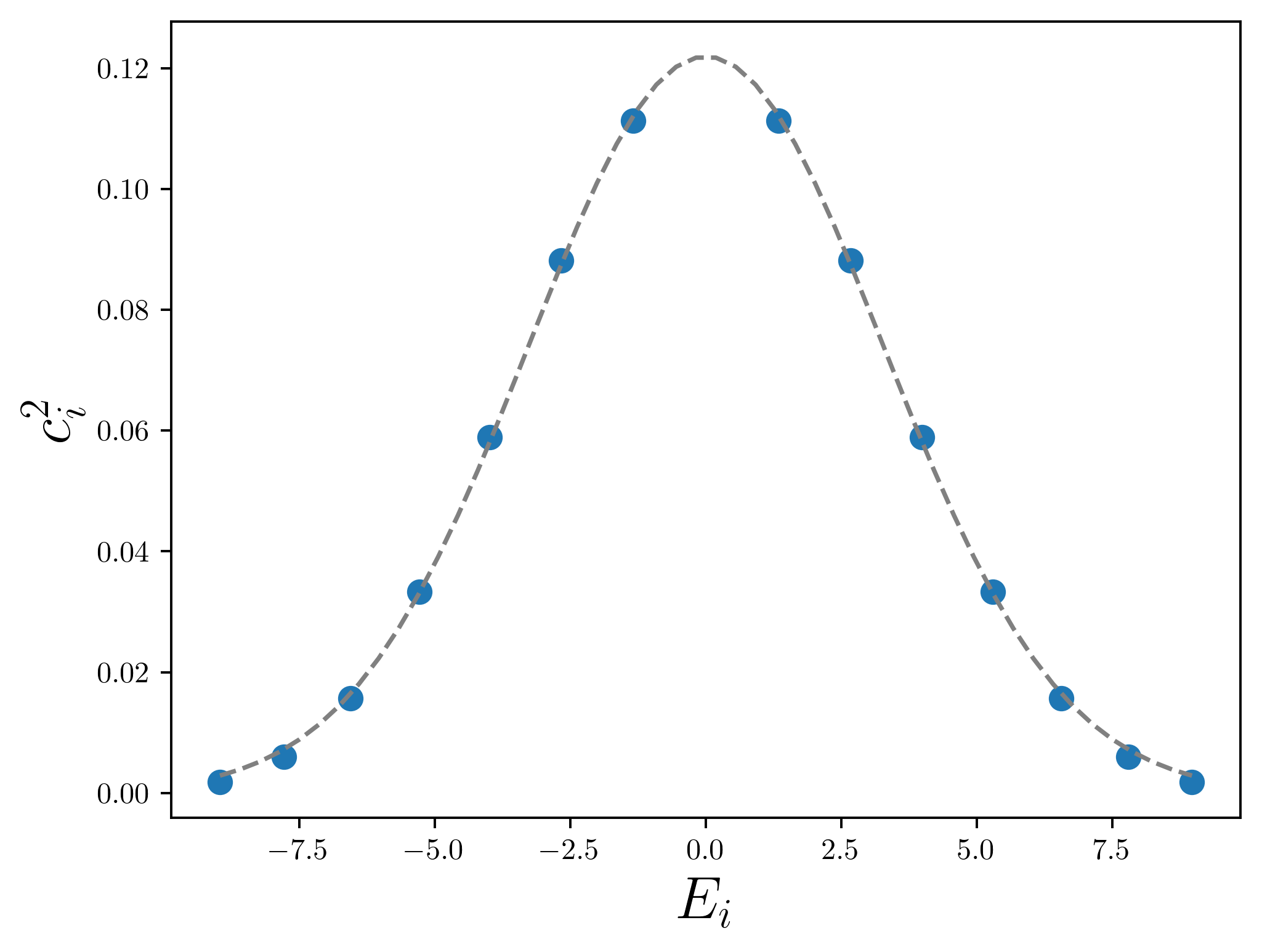}

    \put(0,60){\textbf{(a)}}

    \put(66.2,51.4){
        \begin{overpic}[width=0.295\linewidth]{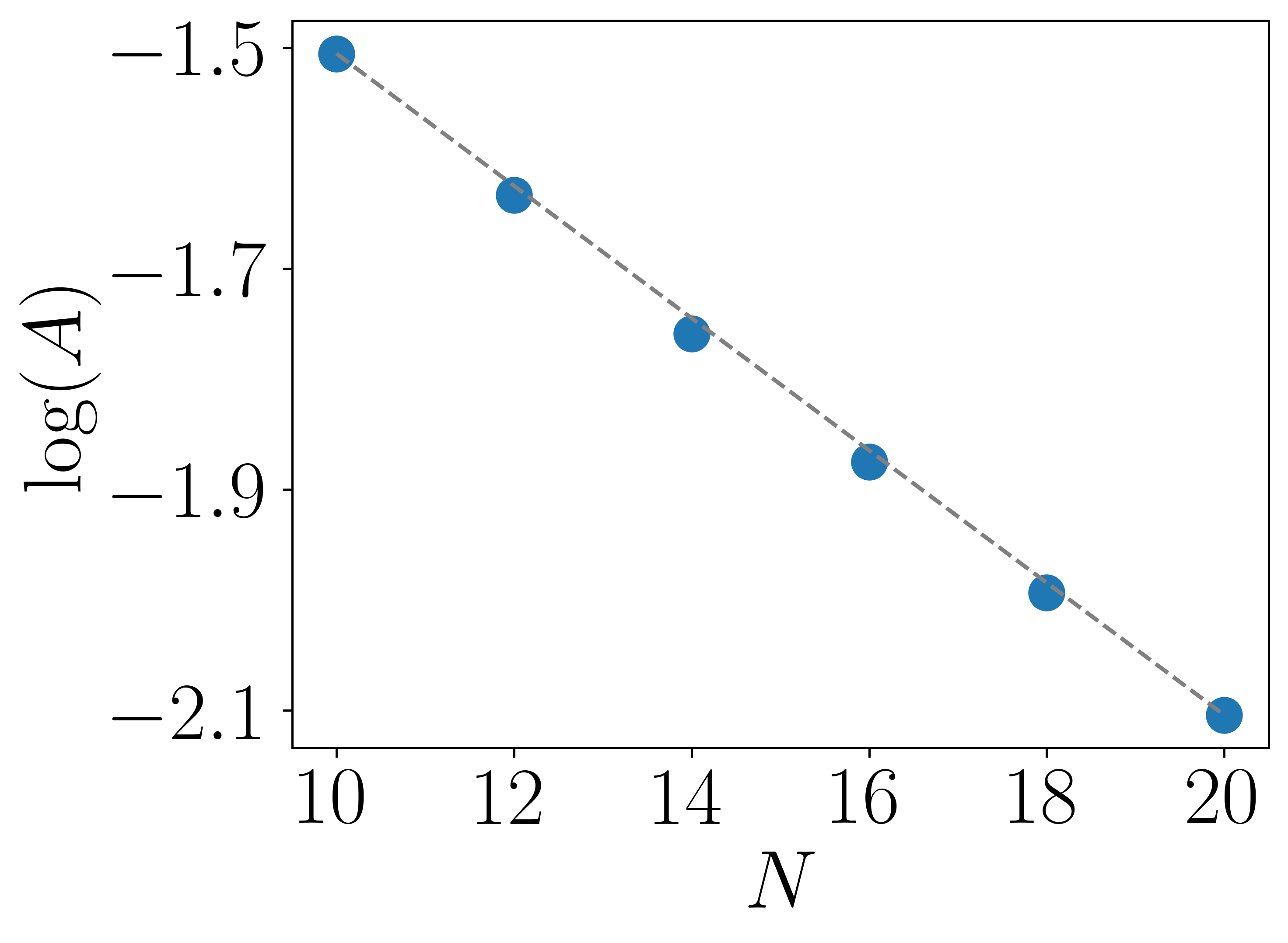}
            \put(50,-12){\textbf{(b)}}
        \end{overpic}
    }

    \put(36,11){
        \begin{overpic}[width=0.345\linewidth]{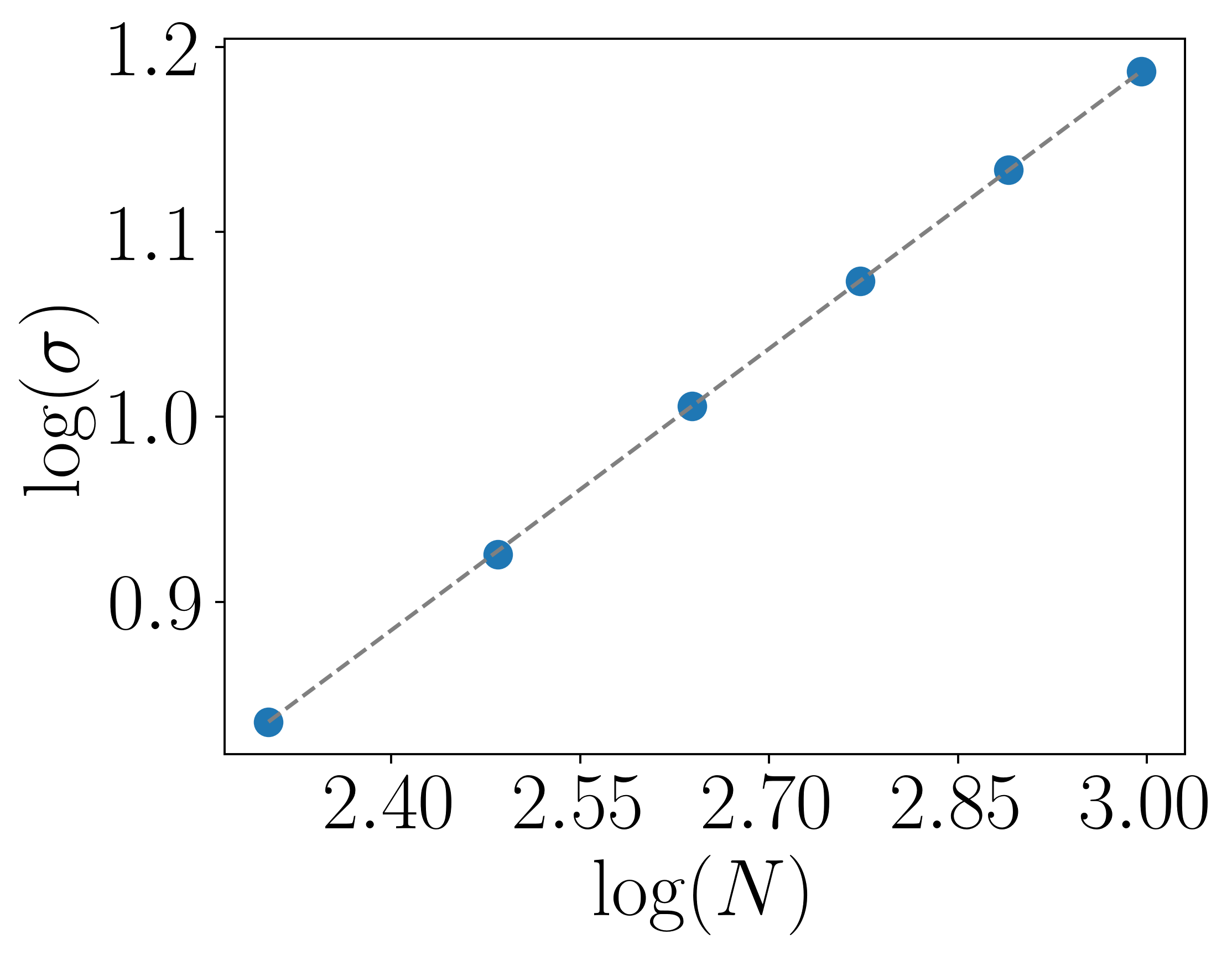}
            \put(-17,25){\textbf{(c)}}
        \end{overpic}
    }

\end{overpic}
\caption{ \textit{Gaussian ansatz.- } In the main panel, we show the initial state overlap with the dominant layer of the MBS spectral tower in a system with $N = 20$ spins. The distribution is Gaussian, with an amplitude that scales exponentially with system size (see inset panel b) and a variance that scales polynomially with $N$ (see inset panel c).
    }
    \label{fig:gaussian}
\end{figure}

\textit{(iii) Gaussian ansatz for overlaps.} The overlap coefficients $c_i = \langle E_i | \mathbb{Z}_2 \rangle$ between scar eigenstates and the initial Néel state are observed to follow a Gaussian distribution (see Fig. \ref{fig:gaussian}),
\begin{equation}
    c_i^2 = A_N \, \exp\left(-\frac{E_i^2}{\sigma_N^2}\right),
    \label{eq:gaussian}
\end{equation}
with amplitude $A_N$ and variance $\sigma_N$. Numerically, we find that $A_N$ decays exponentially with system size, $A_N \approx 0.4 \, e^{-0.06 N}$, while the variance grows polynomially, $\sigma_N \approx 0.71N^{0.5}$.

Under these approximations, the response terms under resonance are given by,
\begin{eqnarray}
\label{eq.approx.response.tot}
    R_{i,i+2}(\omega_{\rm tot},t) &\approx &  R_{\rm tot}(\omega_{\rm tot},t) \\
    &=& \frac{1}{2}\left(\frac{e^{i(2\Delta E-\omega_{\rm tot} )t}-1}{(2\Delta E- \omega_{\rm tot})}-\frac{e^{i(2\Delta E+\omega_{\rm tot} )t}-1}{(2\Delta E+\omega_{\rm tot} )}\right) \nonumber
\end{eqnarray}
for the total magnetization, and
\begin{eqnarray}
\label{eq.approx.response.stg}
R_{i,i+1}(\omega_{\rm stg},t) &\approx  &  R_{\rm stg}(\omega_{\rm stg},t) \\
    &=& \frac{1}{2}\left(\frac{e^{i(\Delta E-\omega_{\rm stg} )t}-1}{(\Delta E- \omega_{\rm stg})}-\frac{e^{i(\Delta E+\omega_{\rm stg} )t}-1}{(\Delta E+\omega_{\rm stg} )}\right) \nonumber
\end{eqnarray}
for the staggered magnetization. Similarly, the AC field operators truncated on their respective subspaces are,
\begin{eqnarray}
 \hat{O}_{\rm tot} &\approx & \sum_i \left( O_{i,i+2}|E_i\rangle\langle E_{i+2}|+O_{i+2,i}|E_{i+2}\rangle\langle E_i|\right) \nonumber \\
    \hat{O}_{\rm stg} &=& \sum_i \left( O_{i,i+1}|E_i\rangle\langle E_{i+1} |+O_{i+1,i}|E_{i+1}\rangle\langle E_i|\right) \nonumber \\
    & &
    \label{eq.approx.connectivities}
\end{eqnarray}

Thus, the QFI can be written as
\begin{equation}
    F_h(t) = |R(\omega,t)|^2 \left[ \sum_i \bra{\psi_0}  \hat{O}^\dagger \hat{O} \ket{\psi_0}-|\bra{\psi_0}  \hat{O} \ket{\psi_0}|^2 \right].
\end{equation}
with response terms and connectivities as given by Eqs.\eqref{eq.approx.response.tot}-\eqref{eq.approx.connectivities}.
Expanding the initial state in the scar eigenbasis $|\psi(0)\rangle = |\mathbb{Z}_2\rangle = \sum_i c_i |E_i\rangle$ reduces the QFI to,
\begin{equation}
    F_h(t) = |R(\omega,t)|^2 \left( \sum_j |d_{j,m}|^2 - \left|\sum_i c_i^* d_{i,m}\right|^2 \right),
    \label{eq:QFI_compact}
\end{equation}
where $d_{i,m}=c_{i+m}O_{i,i+m} + c_{i-m}O_{i,i-m}$ with $m=1$ for staggered magnetization and $m=2$ for the homogeneous  one. Using now explicitly the Gaussian  ansatz for the overlapping terms and  Eqs.\eqref{eq:scaling_stg}-\eqref{eq:scaling_tot} yields
\begin{equation}
\begin{aligned}
&F_h(t) =|R(\omega,t)|^2 (k N^\gamma)^2 0.4 \, e^{-0.06 N} \Bigg[\\
& \sum_j \left(
e^{-\frac{ \left( \Delta E (j+m)\right)^2}{N}}
+ e^{-\frac{\left( \Delta E (j-m)\right)^2}{N}}
\right)^2  - 0.4 \, e^{-0.06 N}  \\
&  \left|\sum_j e^{-\frac{\left( j \Delta E \right)^2}{N}}
\left(
e^{-\frac{\left( \Delta E (j+m)\right)^2}{N}}
+ e^{-\frac{\left( \Delta E (j-m)\right)^2}{N}} \right)\right|^2
\Bigg]
\end{aligned}
\label{eq.qfi.approx}
\end{equation}
 with $k$ and $\gamma$ taking the values $k_{\text{stg/tot}}$ and $\gamma_{\text{stg/tot}}$ from Eqs.~\eqref{eq:scaling_stg}--\eqref{eq:scaling_tot}, and $m=1,2$ indexing the type (staggered/homogeneous) of the AC field.

The above equation predicts an initial quadratic growth of the QFI in time, encoded in the factor $|R_{\rm stg(tot)}(\omega_{\rm stg(tot)},t)|^2 \propto t^2$. This follows directly from the resonance condition and assumption (ii), which neglects finite-size spectral corrections. This behavior is consistent with the numerical results shown in Fig.~\ref{fig:overtime_resonance}, where a rough quadratic-in-time plateau emerges, with its duration extending as system size increases.
We recall, however, that this approximation should be valid for times not too long compared to the effect of such corrections.

Second, we can extract the scaling with the number of spins. Apart from the explicit polynomial factor $(k N^\gamma)^2$, the dominant $N$-dependence enters through the amplitude $A_N \sim e^{-0.06 N}$. Consequently, the overall scaling behaves as
\begin{equation}
    F_h(t)/t^2 \propto N^{2\gamma} e^{-0.06 N}.
\end{equation}
For small system sizes, the polynomial term $N^{2\gamma}$ leads to a superlinear growth of the QFI. However, for macroscopic $N$, the exponential suppression dominates, predicting an eventual decay of the sensor performance with system size. This counterintuitive result - that larger systems may not yield better sensors under this resonant driving protocol - stems from the exponentially small overlap of the initial state with the highest energy scar manifold. In this way, distinct possibilities follow from this analytical result. First, the sensor's performance may indeed deteriorate for very large $N$, as predicted by the exponential suppression in the overlap term. Alternatively, the approximation itself may become inadequate in the macroscopic limit. The latter scenario is plausible given that our model considers only the highest tower of scars. In reality, multiple scar towers emerge in the spectrum and become increasingly important for describing the system's full dynamics as $N$ increases. The neglected contributions from these additional towers could, in principle, compensate for or alter the exponential decay captured by the single-tower approximation.
Therefore, developing an analytical framework that incorporates multiple scar towers represents a compelling avenue for future work.

\section{Conclusion}

In this work, we have investigated the metrological potential of MBS systems as quantum sensors for weak AC fields. Focusing on the PXP model as a paradigmatic scarred Hamiltonian, we analyzed how to harness the non-thermal tower of eigenstates and constrained dynamics to enhance parameter estimation, as quantified by the QFI.
Our analysis demonstrates that two central ingredients govern the sensing performance: (i) the resonant structure of the scar tower, characterized by an approximately uniform energy spacing $\Delta E$, and (ii) the connectivity induced by the probe operator across the tower. By tuning the driving frequency to match integer multiples of the scar gap, $\omega = m \Delta E$, we showed that it is possible to activate the collective resonant processes involving a macroscopic number of scar eigenstates. This mechanism leads to an enhanced coherent response encoded in the Heisenberg signal operator, resulting in quadratic-in-time growth of the QFI over a sizable time window.
We have further demonstrated that the spatial profile of the perturbation plays a decisive role. In particular, the staggered magnetization operator exhibits stronger, more favorable connectivity scaling along the tower than the homogeneous magnetization operator does. This translates into superior metrological performance for staggered driving, as confirmed by both frequency-scanning analysis and finite-size scaling of the dominant matrix elements. In the resonant regime, the duration of the favorable scaling regime increases with system size, highlighting the potential of scar-based sensors to exploit many-body resources in a controlled and robust manner.

A key conceptual message emerging from our work is that maximal entanglement is not the only viable route toward enhanced quantum sensing. While Heisenberg-limited probes such as GHZ states can, in principle, achieve $F_h \propto N^2 t^2$, thermalization severely constrains their performance, thereby reducing the optimal interrogation time as the system size increases. In contrast, weakly thermalizing scar systems offer extended coherent dynamics and correlation-enhanced scaling, providing a competitive — and potentially superior — strategy in realistic scenarios where thermalization limits the usable interrogation time. In this sense, robustness and dynamical protection may represent more scalable metrological resources than extreme multipartite entanglement.

To gain analytical insight, we also developed a single-tower approximation under resonant driving. Within this framework, we derived a compact expression for the QFI that captures both its quadratic time growth and its scaling with system size. While this approximation predicts an eventual exponential suppression of the QFI as $N$ increases, due to the decreasing overlap of the initial state with a single scar manifold, this behavior likely reflects the limitations of truncating to a single tower. In the full spectrum, multiple scar towers emerge and may collectively contribute in the macroscopic limit. Extending the analytical treatment to incorporate these additional structures constitutes an important direction for future work.

Overall, our results establish quantum many-body scars as a promising platform for quantum-enhanced sensing. By exploiting their structured spectrum, constrained dynamics, and anomalously slow thermalization, scar-based sensors offer a novel paradigm to convert non-ergodic many-body dynamics into metrological gain. We anticipate that these ideas can potentially be generalized beyond the PXP model to other scarred systems and weakly thermalizing phases, paving the way toward robust many-body quantum sensors operating beyond standard metrological strategies.

\section{ACKNOWLEDGMENTS}

F.I. acknowledges financial support from the Brazilian funding agencies CAPES, CNPQ, and FAPERJ (Grants N\textsuperscript{o} $151064/2022$-$9$,  N\textsuperscript{o}E-$26/201.365/2022$, N\textsuperscript{o}E-$26/204.340/2025$) and from the Serrapilheira Institute (Grant N\textsuperscript{o} $2211$-$42166$, Serra). A.T. acknowledges financial support from the Brazilian funding agency CAPES. TRO acknowledges funding from the Air Force Office of Scientific Research under Grant No. FA9550-23-1-0092. M.F. acknowledges funding from the Air Force Office of Scientific Research and the Serrapilheira Institute.  

\bibliography{biblio}

\end{document}